\documentclass[aps,prl,twocolumn,nofootinbib,showpacs,superscriptaddress,floatfix]{revtex4-1}  
\usepackage{graphicx}  
\usepackage{dcolumn}   
\usepackage{bm}        
\usepackage{amssymb}   
\usepackage{amsmath}
\usepackage{comment}
\usepackage{graphicx}
\usepackage{slashed}
\usepackage{braket}
\usepackage{color}

\begin{document}

\author{Niklas M\"{u}ller}
\email{n.mueller@thphys.uni-heidelberg.de}
\affiliation{Institut f\"{u}r Theoretische Physik, Universit\"{a}t Heidelberg, Philosophenweg 16, 69120 Heidelberg, Germany}
\author{S\"{o}ren Schlichting}
\email{sschlichting@bnl.gov}
\affiliation{Physics Department, Brookhaven National Laboratory, Bldg. 510A, Upton, NY 11973, USA}
\author{Sayantan Sharma}
\email{sayantans@bnl.gov}
\affiliation{Physics Department, Brookhaven National Laboratory, Bldg. 510A, Upton, NY 11973, USA}

\title{Chiral magnetic effect and anomalous transport from real-time lattice simulations}
\date{\today}
\begin{abstract} 
We present a first-principle study of anomaly induced transport phenomena by performing real-time lattice 
simulations with dynamical fermions coupled simultaneously to non-Abelian $SU(N_c)$ and Abelian $U(1)$ gauge 
fields. Investigating the behavior of vector and axial currents during a sphaleron transition in the presence 
of an external magnetic field, we demonstrate how the interplay of the Chiral magnetic (CME) and Chiral 
separation effect (CSE) lead to the formation of a propagating wave. We further analyze the dependence of the 
magnitude of the induced vector current and the propagation of the wave on the amount of explicit chiral 
symmetry breaking due to finite quark mass.
\end{abstract}
\maketitle

Novel transport phenomena associated with the presence of chiral fermions have generated 
enormous excitement across the physics community. Soon after it was realized that local imbalances of the axial 
charge $j^{0}_{a}$ can lead to a series of new transport phenomena \cite{Vilenkin:1980fu,Fukushima:2008xe}, 
otherwise forbidden by discrete symmetries, a comprehensive search of possible manifestations 
has started to take place across the most diverse range of energy scales, ranging from heavy-ion collisions 
in high-energy QCD \cite{Kharzeev:2007jp, Kharzeev:2010gr} to condensed matter experiments with Dirac and 
Weyl semi-metals \cite{Li:2014bha}. The most prominent example of such anomalous transport phenomena 
is the Chiral Magnetic effect (CME) \cite{Fukushima:2008xe}, whereby an external magnetic field $\vec{B}$ 
can induce an electro-magnetic (vector) current along its direction, $\vec{j}_{v} \propto j^{0}_{a} \vec{B}$. 
However, there are in fact a variety of transport phenomena that occur due to non-trivial interplay of axial 
and vector charges and we refer to \cite{Kharzeev:2015kna} for a comprehensive review.

While in Abelian gauge theories such as QED, the only source of an axial charge imbalance is due to parallely 
oriented electric and magnetic fields,  in non-Abelian gauge theories like  QCD there 
is an additional possibility to generate a local imbalance of axial charges via topological transitions 
\cite{Fukushima:2008xe}. In the presence of a sufficiently strong magnetic field, the CME and its 
associated phenomena can therefore act as a unique messenger of the real-time dynamics of topological 
transitions in QCD.

It was proposed early on \cite{Kharzeev:2007jp, Kharzeev:2010gr} that aforementioned effects could be observed in 
heavy-ion collision experiments at the Relativistic Heavy Ion Collider (RHIC), where strong transient magnetic fields 
are generated at early times due to the charged nuclei moving almost at the speed of light. Experimentally intriguing 
hints of the CME have been observed \cite{Abelev:2009ac, Abelev:2012pa,Adamczyk:2014mzf}, however the situation remains 
controversial as the measurements are also subject to large background uncertainties \cite{Kharzeev:2015znc}.  
Moreover, since the lifetime of the magnetic field is expected to be short $< 1$ fm/c \cite{Skokov:2009qp, Tuchin:2013apa},  
the theoretical description of these effects is also challenging as it requires a first-principle description of 
far-from-equilibrium dynamics at very early-times \cite{Mace:2016svc}. 

So far the dynamics of the CME has been studied primarily in near-equilibrium situations both microscopically using perturbation 
theory \cite{Fukushima:2008xe}, euclidean lattice techniques~\cite{Buividovich:2009wi} or holography \cite{Yee:2009vw,Amado:2011zx, Lin:2013sga} as well as macroscopically using anomalous 
extensions of relativistic hydrodynamics \cite{Newman:2005hd,Son:2009tf,Hirono:2014oda,Yin:2015fca}. Chiral kinetic theory 
\cite{Stephanov:2012ki,Gao:2012ix,Son:2012zy,Gorbar:2016qfh} is another theoretical approach which has been developed recently 
and in principle extends beyond near-equilibrium situations, but its study has been limited to Abelian gauge theories so far. 
In this letter we establish a new first-principles technique based on real-time lattice gauge theory simulations with dynamical 
fermions. Our approach is specifically devised to study anomalous transport phenomena in far-from-equilibrium situations, 
encountered e.g. during the early stages of high-energy heavy-ion collisions.

Classical-statistical lattice gauge theory simulations have been an essential tool to study the non-perturbative early-time 
dynamics immediately after the collision of heavy nuclei \cite{Berges:2013eia,Berges:2013fga,Kurkela:2012hp,Berges:2012ev}.
While significant progress has been made in understanding the dynamics of gauge fields, including for the first time the dynamics of topological transitions at early times \cite{Mace:2016svc}, 
the non-equilibrium dynamics of fermions \cite{Aarts:1998td,Borsanyi:2008eu} is only starting to be explored with first attempts made 
to elucidate the fermion production mechanism in non-Abelian \cite{Gelis:2005pb,Saffin:2011kn,Gelfand:2016prm,Tanji:2016dka} and 
Abelian gauge theories \cite{Hebenstreit:2013qxa,Kasper:2014uaa,Mueller:2016aao}. However, to explore the real-time dynamics 
of anomaly induced charge transport, it is essential to include dynamical light quarks, 
coupled simultaneously to the non-Abelian (QCD) and Abelian (QED) gauge fields, into the non-equilibrium description. 
In this letter we report the first real-time lattice study of this nature, exploring the dynamics of quarks during 
and after a sphaleron transition in the presence of an external magnetic field. We focus on the evolution of 
vector and axial currents and for simplicity treat both the non-Abelian and Abelian gauge fields as classical 
backgrounds ignoring the back-reaction of fermions.

Our real-time lattice simulations are explained in more detail in the next section, where we also introduce the 
relevant observables. Subsequently, we present simulation results for the dynamical evolution of axial and vector 
currents,  and analyze the effects of explicit chiral symmetry breaking due to the quark mass. We conclude with a 
brief summary of our findings and perspectives for future applications of this framework.

\textbf{Simulation technique} 
We employ the Hamiltonian lattice formalism and discretize the theory on a spatial lattice of 
dimensions $N_x\times N_y\times N_z$ and spacing $a_s$, where the non-Abelian $SU(N_c)$ and 
Abelian $U(1)$ fields in temporal ($A_{0}=0$) gauge are represented in terms of the lattice 
link variables $U_{x,j} \in SU(N_c) \times U(1)$ with $x$ labeling the lattice site and $j=1,2,3$ 
the Lorentz index. We discretize the fermions using a tree-level $\mathcal O(a^n)$ improved version 
of the Wilson Hamiltonian
\begin{eqnarray}
 \label{eq:FermionicHamiltonian}
\hat{H} = a_{s}^3 \sum_{x} \hat{\psi}^{\dagger}_{x}  \gamma^{0} (-i\slashed{D}^{s}_{W}+ m) \hat{\psi}_x ~,
 \end{eqnarray}
 where
 \begin{eqnarray}
 \slashed{D}^{s}_{W}\hat \psi_x=\frac{1}{2a_s} \sum_{n,j} C_{n} 
  \Big[ \Big(\gamma^{j}-in r_{w} \Big) U_{x,nj} \hat{\psi}_{x+nj} \nonumber \\
 +2 i n r_{w} \hat{\psi}_{x} - \Big(\gamma^{j}+i n r_{w} \Big)U_{x,-nj} \hat{\psi}_{x-nj} \Big]~,
\end{eqnarray}
with the coefficients $C_{n}$ fixed in order to explicitly cancel the $\mathcal{O}(a^n)$ terms. While we 
found the next-to-leading order improvement with $C_{1}=4/3,C_{2}=-1/6$ to be important, higher
order improvement was also checked and found not to provide a significant advantage \cite{mmss}. 
Evolution equations for the fermion operators, are derived from the lattice Hamiltonian and take the usual form
\begin{eqnarray}
\label{eq:DiracEquation}
i \gamma^{0} \partial_t \hat \psi=(-i\slashed{D}^{s}_{W}+m)\hat \psi \;.
\end{eqnarray} 
which we solve numerically using a modefunction expansion~\cite{Aarts:1998td}. Since for a classical gauge field 
configuration Eq.~(\ref{eq:DiracEquation}) is linear in the fermion operator $\hat{\psi}$, one 
can expand its solution in the basis of creation ($\hat{b}_{\lambda}(0)$) and annihilation 
operators ($\hat{d}^{\dagger}_{\lambda}(0)$) of fermions and anti-fermions at initial time, $t=0$,
\begin{eqnarray}
\hat{\psi}_x(t)=\frac{1}{\sqrt{V}} \sum\limits_{\lambda}\left( \hat{b}_{\lambda}(0)\phi^u_{\lambda}(t,x) +
\hat{d}^{\dagger}_{\lambda}(0)\phi^v_{\lambda}(t,x)  \right)\;,
\end{eqnarray}
where $\lambda$ labels the eigenmodes of the Hamiltonian. By construction the time dependence is then 
entirely reflected by the wave-functions $\phi^{u/v}_{\lambda}(t,x)$, which we calculate by diagonalizing 
the Hamiltonian at initial time and subsequently solving the Dirac equation for each mode-function using a 
leap-frog scheme with a time step $a_{t}=0.02a_s$. When computing physical observables, operator expectation 
values are evaluated according to the density matrix of the initial state; here we simply consider an 
initial vacuum for which $<[b^{\dagger}_{\lambda},b_{\lambda'}]>=2 (n_\lambda^{u}-1/2)
\delta_{\lambda,\lambda'}$ and $<[d_{\lambda},d_{\lambda'}^{\dagger}]>=-2(n_\lambda^{v}-1/2) \delta_{\lambda,\lambda'}$ 
with $n_{\lambda}^{u/v}=0$ give the only non-vanishing contributions to fermion bilinears~\cite{Kasper:2014uaa}.

\textit{Gauge links}: Since the computational cost of this calculation is significant, we consider for simplicity 
a sphaleron transition in two-color QCD, noting however that the extension to $SU(3)$ is straightforward and will not 
change the essential features of the problem. Specifically, we construct a topologically non-trivial map 
$G:T^3_\text{lattice}\rightarrow S^3_{SU(2)}$, which extends over a characteristic scale 
$r_{\rm{sph}}$ and has winding equal to unity. Our $SU(2)$ gauge links are then constructed \cite{mmss} by connecting 
the topologically distinct vacua $U^{SU(2)}_{x,i}(t=0)=\mathbf{1}$ and $U^{SU(2)}_{x,i}(t\geq t_{{\rm sph}})=G_xG^{\dagger}_{x+i}$ 
along a trajectory with constant electric field $E^{SU(2)}_{x,i}$ over the time scale $t\in[0,t_{{\rm sph}}]$. Concerning 
the $U(1)$ gauge links, we employ a constant magnetic field pointing in the $z$ direction, i.e. $\vec{B}=B\hat{z}$, which 
is implemented on the lattice following Ref. \cite{AlHashimi:2008hr,Bali:2011qj}. Back coupling of fermions to the evolution 
of the gauge fields is not considered in this work.

\begin{figure*}[t!]
\centering
\includegraphics[width=0.9\textwidth]{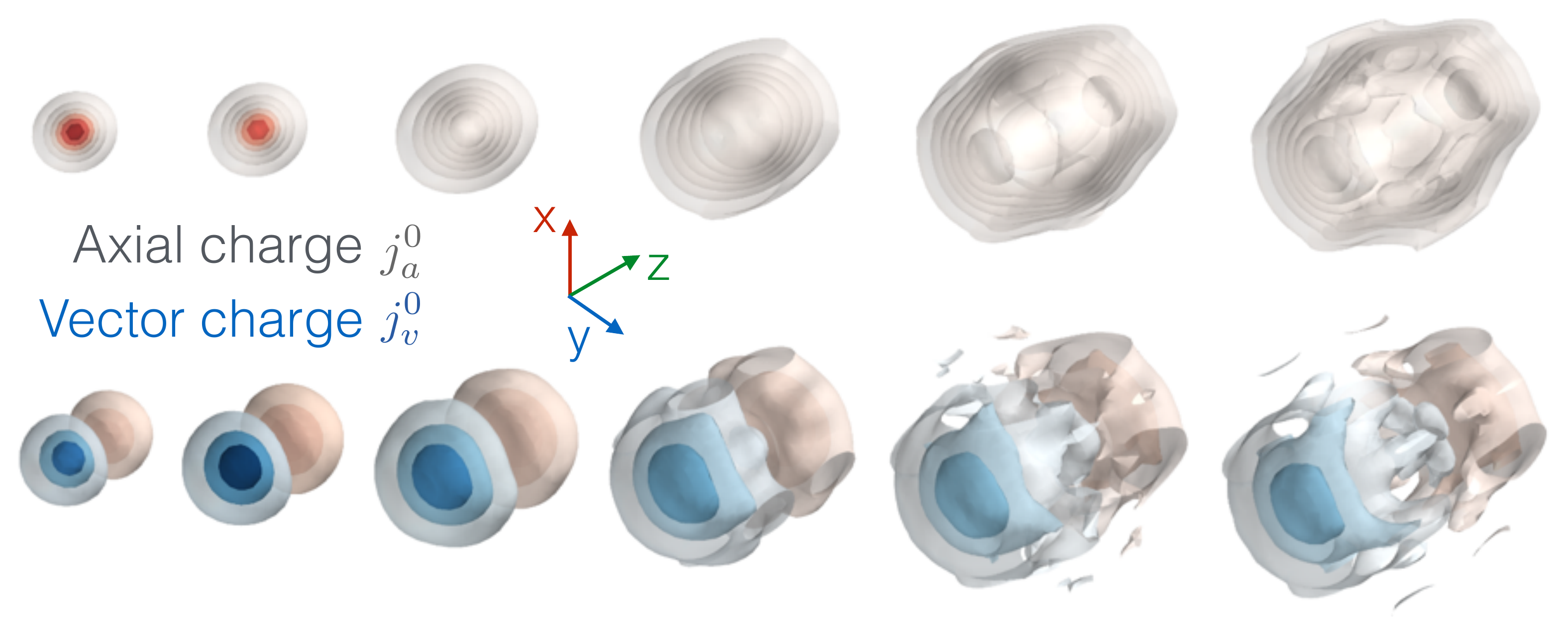}
\caption{ Illustration of real-time dynamics of the chiral magnetic wave for light quarks ($mr_{\rm{sph}} \ll 1$ ). 
Contour lines represent the distribution of axial and vector charges at times $t/t_{{\rm{sph}}}=0.6,\;0.9,\;1.1,\;1.3,\;
1.6,\;1.9$ of the evolution. Simulations were performed on a $24\times 24\times 64$ lattice.}
\label{CMW}
\end{figure*}

\textit{Observables:} 
Vector and axial densities are defined in analogy to the continuum as 
$j^{0}_{v}(x)=\langle \hat \psi^\dagger_{x} \hat \psi_{x}\rangle$ and $j^{0}_{a}(x)=\langle \hat \psi^\dagger_{x}  \gamma_{5} 
\hat \psi_{x}\rangle$, whereas 
the spatial components of the currents take the form, 
\begin{eqnarray}
\nonumber
j^{i}_{v}(x)&=& \sum_{n,k=0}^{n-1} \frac{C_{n}}{2} \Big \langle \hat{\psi}^\dagger_{x-ki} \gamma^0 \Big(\gamma^{i}-in r_{w} \Big) 
U_{x-ki,ni}~\hat \psi_{x+(n-k)i}  \\ \nonumber
&+& \hat{\psi}^\dagger_{x+(n-k)i} \gamma^0 \Big(\gamma^{i}+in r_{w} \Big)U_{x+(n-k)i,-ni}~\psi_{x-ki} \Big  \rangle ~;
\\\nonumber
j^{i}_{a}(x)&=& \sum_{n,k=0}^{n-1} \frac{C_{n}}{2} \Big \langle \hat{\psi}^\dagger_{x-ki} \gamma^0 \gamma^i \gamma_5~ 
U_{x-ki,ni}~\hat \psi_{x+(n-k)i}  \\
&+& \hat{\psi}^\dagger_{x+(n-k)i} \gamma^0 \gamma^i \gamma_5~ U_{x+(n-k)i,-ni}~\psi_{x-ki} \Big  \rangle~.
\end{eqnarray}
Since the currents are derived from the improved Hamiltonian, by variation with respect to the (Abelian) gauge fields, 
these are by construction improved to the same order, which is crucial for reducing discretization effects in our 
real-time simulations. 

While the vector current is covariantly conserved $\partial_{\mu}j^{\mu}_{v}=0$ as in the continuum, the lattice 
definition of the axial current for Wilson fermions satisfies the anomaly equation  
$\partial_{\mu}j^{\mu}_{a}(x)=2 m  \eta_{a}(x)  + r_{w} \langle W(x) \rangle$ where $\eta_{a}(x)\equiv \langle
\hat{\psi}_{x}^\dagger i \gamma^0 \gamma_{5} \psi_{x} \rangle$ denotes the pseudo-scalar density and $r_w \langle W(x) \rangle$ is the 
contribution of the Wilson term to the anomaly equation.  Since the Wilson term has a non-trivial continuum 
limit \cite{Karsten:1980wd,Tanji:2016dka}  $r_{w} \langle W(x)\rangle \to -\frac{g^2}{8\pi^2}  \text{Tr} 
F_{\mu\nu}(x)\tilde{F}^{\mu\nu}(x)$, the usual expression
\begin{equation}
\partial_{\mu}j^{\mu}_{a}(x)=2 m \eta_{a}(x) -\frac{g^2}{8\pi^2}  \text{Tr}  F_{\mu\nu}(x)\tilde{F}^{\mu\nu}(x) 
 \label{eq:anomalycontinuum}
\end{equation}
is recovered in the continuum and we have carefully monitored the residual cutoff effects in our simulations \cite{mmss}.

Since the spatial size of the sphaleron $r_{\rm{sph}}$, is the only relevant scale in our simulations we will 
express all physical quantities in units of $r_{\rm{sph}}$. As sphalerons are non-perturbative infrared objects with a 
characteristic size of the magnetic screening length (see e.g.~\cite{Bodeker:1998hm,Moore:2010jd,Mace:2016svc}) 
a conversion to physical units can be achieved by assigning a value of about $200 - 500$ MeV to $r^{-1}_{\rm{sph}}$. If not stated otherwise 
we use a lattice spacing of $r_{\rm{sph}}/a=6$, the duration of the sphaleron transition is chosen as
$t_{\rm{sph}}=3/2 r_{\rm{sph}}$ (corresponding to $\sim~0.6-1.5$ fm/c) and the magnetic field strengths considered in 
this work is $qB=3.5r_{\rm{sph}}^{-2}$ (corresponding to a few $m_\pi^2$).


\textbf{Non-equilibrium dynamics of axial and vector charges}
We will now analyze the dynamics of the axial and vector charges during and after a sphaleron transition, 
and first focus on the anomalous transport of light quarks with $mr_{\rm{sph}}\ll 1$, where 
dissipative effects due to a finite quark mass can be neglected over the time scale of a sphaleron transition.
Our results for the time evolution of the axial and vector densities $j^{0}_{v/a}(t,x)$ in the presence of an external 
magnetic field are compactly summarized in Fig.~\ref{CMW} where we show contour lines of the 
distributions at various stages of the time evolution.

We observe that during the sphaleron transition, a local imbalance of axial charge $j^{0}_{a}$ is generated according to the axial 
anomaly; at the same time the chiral magnetic effect induces a vector current $j^{z}_{v}$ with a similar profile in coordinate space. 
Conservation of the vector current $\partial_{\mu}j^{\mu}_{v}=0$, implies that longitudinal gradients of the current 
$\partial_{z}j^{z}_{v}$, lead to separation of electric charges along the direction of $\vec B$; over time electric charge 
accumulates at the edges of the sphaleron, resulting in a dipole like structure of the vector charge density $j^{0}_{v}$ 
observed e.g. at $t/t_{\rm{sph}}=0.6,\;0.9$ in Fig.~\ref{CMW}.

Since the local imbalance of vector charge $j^{0}_{v}$ in turn induces an axial current $\vec{j}_{a} \propto j^{0}_{v} \vec{B} $ 
due to the chiral separation effect (CSE) \cite{Son:2004tq}, the combination of CME and CSE ultimately leads to the formation 
of a chiral magnetic wave \cite{Newman:2005hd,Kharzeev:2010gd}. We observe from Fig.~\ref{CMW}, that the chiral magnetic wave manifests 
itself as the propagation of a soliton-like wave-packet associated with the non-dissipative transport of axial and vector 
charges along the direction of magnetic field. We note that this is the first time that the emergence of such a collective 
excitation is confirmed in non-perturbative real-time lattice gauge theory simulations.   

\textit{Chiral magnetic wave:} The dynamics of the chiral magnetic wave can be further investigated by integrating out the 
transverse coordinates to study the propagation of the wave-packet along the longitudinal direction.  This allows us to 
compare the results of our microscopic simulations with a macroscopic description within the framework of anomalous 
hydrodynamics in a straightforward way. In anomalous hydrodynamics \cite{Newman:2005hd,Son:2009tf,Hirono:2014oda,Yin:2015fca}, 
the coupled dynamics of axial and vector charges 
is described in terms of conservation laws and the constitutive relations of the currents, which to leading 
order in gradients and in presence of an external magnetic field $B^{\mu}=(0,0,0,B)$ take the form \cite{Son:2009tf}
\begin{eqnarray}
j^{\mu}_{v,a}=n_{v,a} u^{\mu} + D_{v,a} \triangledown^{\mu} n_{v,a} +\sigma^B_{v,a} B^{\mu}\;.
\end{eqnarray}
Specifically, for a system of non-interacting fermions, the diffusion constant $D_{v,a}$, vanishes and the anomalous 
conductivities are simply given by $\sigma^B_{v,a}=n_{a,v}/B$ when the magnetic field strength is sufficiently large  
$B\gg r_{\rm{sph}}^{-2},m^2$ \cite{Fukushima:2008xe}.   In the local rest-frame $u^{\mu}=(1,0,0,0)$, the anomalous hydrodynamic 
equations of motion for the integrated quantities $j^{0,z}_{v,a}(t,z)=\int d^2 x_{\bot}~j^{0,z}_{v,a}(t,x_{\bot},z)$  then take 
the form,
\begin{eqnarray}
\partial_{t} \begin{pmatrix} j^{0}_{v}(t,z) \\ j^{0}_{a}(t,z) \end{pmatrix}= -\partial_{z} \begin{pmatrix} j^{0}_{a}(t,z) 
\\ j^{0}_{v}(t,z) \end{pmatrix} +  \begin{pmatrix} 0\\ S(t,z) \end{pmatrix}\;
\end{eqnarray}
with the sphaleron induced source term given as $S(t,z)= -\frac{g^2}{8 \pi^2} \int d^2 x_{\bot} \text{Tr}~F^{\mu\nu}
\tilde{F}_{\mu\nu}$. The solutions for the vector and axial currents can be then constructed easily,
\begin{eqnarray}
&& j^{0}_{v,a}(t>t_{\rm{sph}},z) = \\
&&\quad \frac{1}{2} \int_{0}^{t_{\rm{sph}}} dt' ~\Big[ S\big(t',z-c (t-t')\big)  \mp S\big(t',z+c (t-t')\big) \Big]   
\nonumber
\end{eqnarray}
which correspond to wave-packets moving along $\pm z$ directions at the speed of light, $c$.

\begin{figure}[t!]
\centering
\includegraphics[width=0.94\columnwidth]{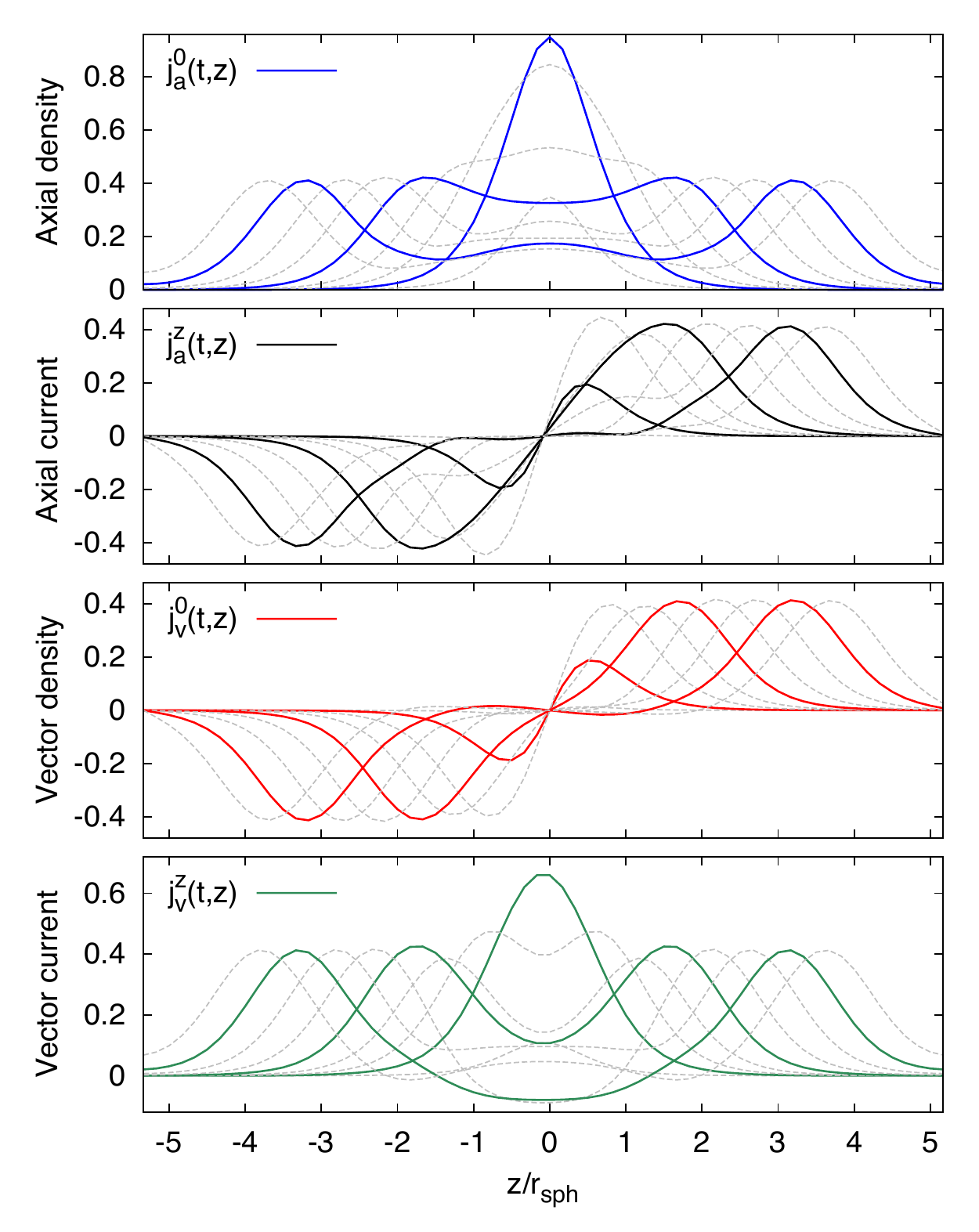}
\caption{ Longitudinal profiles of the axial and vector densities and currents measured in units of $r_{\rm{sph}}^{-1}$ 
at different times of the evolution. Solid curves correspond to times $t/t_{\rm{sph}}=0.67\;,1.67,\;2.67$ (inside-out); 
dashed lines correspond to intermediate time steps.}
\label{Currents}
\end{figure}

Our lattice results for the dynamics of (transversely) integrated currents $j^{0,z}_{v,a}(t,z)$ are presented in 
Fig.~\ref{Currents}, showing the longitudinal profiles at different times of the evolution. One clearly observes that 
subsequent to a sphaleron transition, the axial charge is transported away from the center of the sphaleron generating 
a positive (negative) vector charge propagating along the $\pm z$ directions respectively. We find that once an axial 
charge imbalance is generated from the microscopic QCD dynamics, qualitative features of the transport of vector and 
axial charges such as the overall magnitude and symmetry properties of currents as well as the wave velocity are indeed 
well described by our simple analysis in anomalous hydrodynamics. However, our microscopic description also reveals 
minor deviations, e.g., a smaller fraction of the axial charge remains localized at the center of the sphaleron 
throughout the evolution.

\begin{figure}[t!]
\centering
\includegraphics[width=\columnwidth]{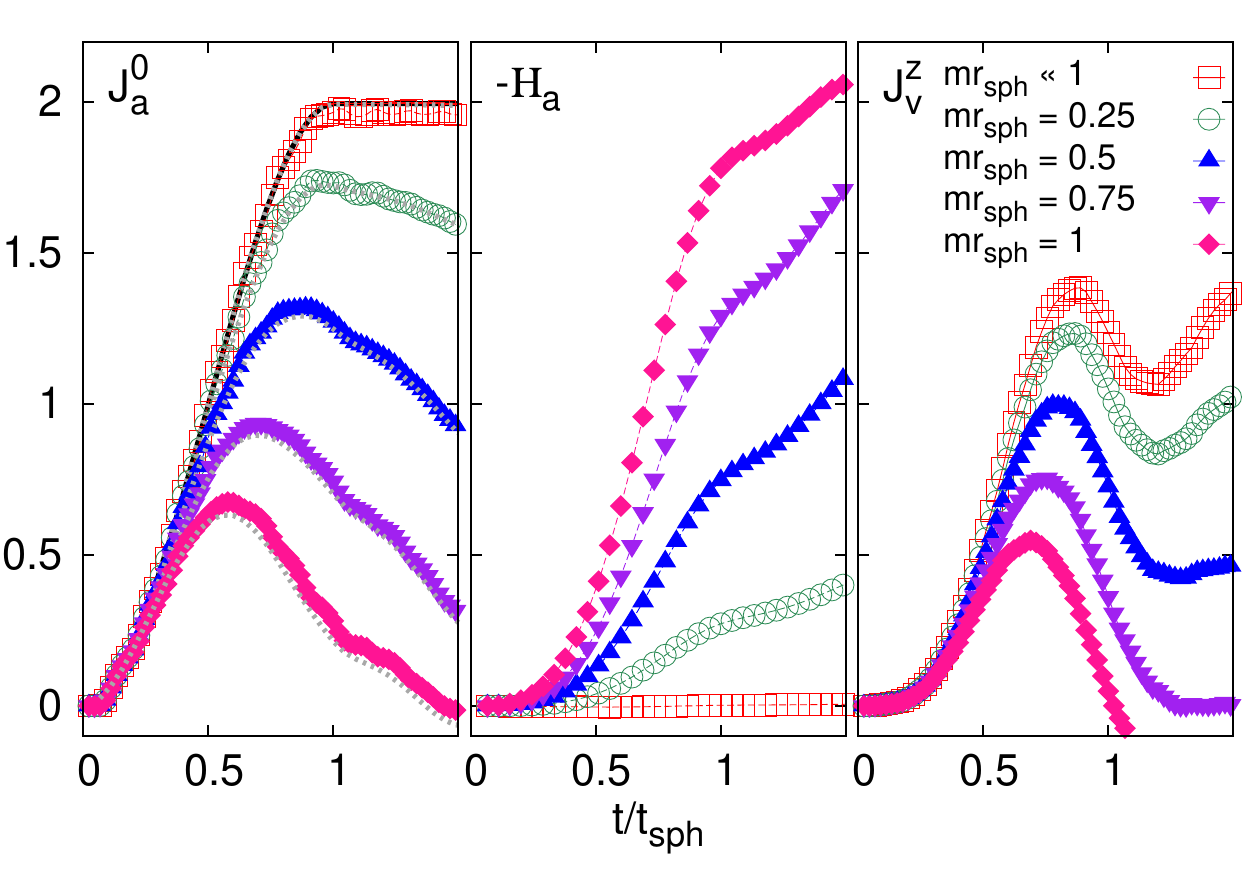}
\caption{ Evolution of the axial charge $J^{0}_{a}$ (left), 
pseudo-scalar condensate $H_a$ (center) and vector current $J^{z}_{v}$ (right) for different values of the fermion mass 
$m r_{\rm{sph}}=3\cdot 10^{-3},0.25,0.5,0.75,1.0$. Comparison with the gray lines in the left panel demonstrates that 
the axial anomaly relation (\ref{eq:anomalycontinuum}) is satisfied in all cases.}
\label{QuarkMassDependence}
\end{figure}

\textit{Quark mass dependence: } 
So far we have investigated the real-time dynamics of anomaly induced transport for light quarks ($mr_{\rm{sph}}\ll1$). We will 
now vary the quark mass to study the effects of explicit non-conservation of axial charge. In Fig.~\ref{QuarkMassDependence} 
we show the time evolution of the net axial charge $J^{0}_{a}(t)=\int d^3x\;j^0_{a}(t,x)$, the pseudo-scalar condensate
$H_a(t)=\int_{0}^{t}dt' \int d^3x\;\eta_{a}(t',x)$ and the net vector current  $J^{z}_{v}(t)$ in the presence of a magnetic field 
$qB=3.5r_{\rm{sph}}^{-2}$ for different values of the quark mass $m r_{\rm{sph}}=3\cdot 10^{-3}, 0.25, 0.5, 0.75, 1$. 
Despite the fact that our lattice simulations accurately reproduce the anomaly relation Eq.~(\ref{eq:anomalycontinuum}) 
in all cases, clear differences emerge for the different scenarios.  With light fermions the evolution of the net axial 
charge $J^0_{a}$ closely follows that of the Chern-Simons number $\Delta N_{CS}(t)= \frac{g^2}{16 \pi^2} \int_{0}^{t} dt' 
\int d^3x~\text{Tr}~F^{\mu\nu} \tilde{F}_{\mu\nu}$ and the sphaleron transition induces two units of net axial charge.  
In contrast, for heavier fermions a significant fraction of the anomaly budget is absorbed by the growth of the pseudo-scalar 
condensate $H_{a}(t)$, which in turn results in a reduction of the axial charge imbalance $J^{0}_{a}$ with increasing quark mass. 
Even for an intermediate quark mass $m r_{\rm{sph}}=1/2$, which is of the order of strange quark mass, the maximal imbalance of axial charges is reduced by a factor of two. 

Similar to the axial charge imbalance, we find that the behavior of the induced vector current $J^{z}_{v}$  also exhibits significant 
changes with increasing quark mass. While for relatively light quarks ($m r_{\rm{sph}}\lesssim 1/4$), the overall magnitude 
of the vector current simply reduces, the changes for the heaviest quark mass $m r_{\rm{sph}}=1$ are more dramatic as the 
current  $J^{z}_{v}$ even reverses its direction immediately after the sphaleron transition. Beyond the time scales shown 
in Fig.~\ref{QuarkMassDependence}, we find that for heavier masses both the net axial charge $J^{0}_{a}(t)$ and vector 
current $J^{z}_{v}(t)$ oscillate in time and we expect that they will eventually decay due to chirality changing 
interactions \cite{Grabowska:2014efa,Manuel:2015zpa}. While at present finite volume effects in our simulations for heavier 
quarks prohibit us to quantitatively access the dynamics at later times, we intend to return to this issue in a future 
publication \cite{mmss,mtb}.

\textbf {Conclusions \& Outlook.} We established a new theoretical method to explore some of the fascinating phenomena of 
anomaly induced transport under far from equilibrium conditions. As a first application, we demonstrated how real-time fermion 
production in the presence of a QCD sphaleron transition leads to a net chirality imbalance. We showed how, in the 
presence of a magnetic field, the chiral magnetic and chiral separation effects lead to the formation of a chiral magnetic 
wave, separating electric charges along the direction of the magnetic field. Many aspects of anomalous transport predicted 
earlier in near-equilibrium situations are beautifully captured within our framework, which allows for the first time to 
study the complex real-time dynamics of these effects from first principles in the underlying quantum field theory. We 
showed that, for sufficiently strong magnetic fields, the transport of the currents could be qualitatively described 
within the framework of anomalous hydrodynamics and demonstrated that for heavier fermions explicit chiral symmetry
breaking due to finite quark masses leads to a significant reduction of the axial charge production and the anomalous 
vector currents.

Our simulations provide a first step towards a realistic description of the dynamics of the chiral magnetic effect and related 
phenomena in relativistic heavy-ion collisions. Since the lifetime of the magnetic field is short, and the rate of 
topological transitions is largest at early times \cite{Mace:2016svc}, a significant part of the effect is expected 
to occur during the first fm/c where a microscopic field theoretical description is inevitable. Of course, to obtain 
quantitative estimates of the chirality imbalance and vector currents at early-times, it will be important to include 
fermionic back-reaction and extend our study to more realistic ensembles of non-equilibrium gauge fields and work is 
in progress along these directions \cite{mmss}. Ultimately, our work could be extended towards providing the initial 
conditions for the anomalous hydrodynamic evolution all the way till hadronization, to understand manifestations of the 
CME in hadronic observables. 

Since our real-time lattice simulations allow to study the complex 3+1D space-time dynamics of anomalous transport phenomena in 
a variety of quantum field theories, we anticipate many further interesting applications beyond high-energy QCD. As an 
example, this method can be easily extended to study anomalous transport properties in strongly correlated electron systems, chiral plasma instabilities \cite{Akamatsu:2013pjd,Buividovich:2015jfa,Akamatsu:2015kau} or fundamental matter produced in collisions of ultra-strong laser pulses \cite{Mueller:2016aao}. 

 \begin{acknowledgments}
We would like to thank J\"{u}rgen Berges, Frithjof Karsch, Dima Kharzeev, Mark Mace, Larry McLerran, Guy D. Moore, Alexander Rothkopf, Naoto Tanji, 
Raju Venugopalan, Ho-Ung Yee and Yi Yin for insightful discussions. SoS and SaS are supported under DOE Contract No. DE-SC0012704. This research used 
resources of the National Energy Research Scientific Computing Center, a DOE Office of Science User Facility supported by the Office 
of Science of the U.S. Department of Energy under Contract No. DE-AC02-05CH11231. Part of this work was performed on the computational 
resource \mbox{ForHLR Phase I} funded by the Ministry of Science, Research and the Arts Baden-W\"urttemberg and DFG 
("Deutsche Forschungsgemeinschaft"). SoS gratefully acknowledges a Goldhaber Distinguished Fellowship from Brookhaven Science 
Associates. NM\ acknowledges support by the Studienstiftung des Deutschen Volkes and by the DFG Collaborative Research Centre SFB 1215 (ISOQUANT). 
\end{acknowledgments}


\end{document}